\documentclass[]{aa}
\usepackage[dvips]{color}
\usepackage{epsfig}
\usepackage{graphicx}

\def\etal{{et~al.~}}
\def\ie{{\em i.e. }}

\def\deg{$^\circ$}
\titlerunning{INTEGRAL/SPI ground calibration}
\authorrunning{Atti\'e \etal}

\begin{document}
%%%%%%%%%%%%%%%%

%%% ----------------------------------------------------------------
%%%       Title
%%% ----------------------------------------------------------------

\title{INTEGRAL/SPI ground calibration}

%%% ----------------------------------------------------------------
%%%       Authors
%%% ----------------------------------------------------------------

\author{
 D. Atti\'e\inst{1} \and
 B. Cordier\inst{1} \and
 M. Gros\inst{1} \and
 Ph. Laurent\inst{1} \and
 S. Schanne\inst{1} \and
 G. Tauzin\inst{1} \and
 P. von Ballmoos\inst{2} \and
 L. Bouchet\inst{2} \and
 P. Jean\inst{2} \and
 J. Kn\"{o}dlseder\inst{2} \and
 P. Mandrou\inst{2} \and
 Ph. Paul\inst{2} \and
 J.-P. Roques\inst{2} \and
 G. Skinner\inst{2} \and
 G. Vedrenne\inst{2} \and
 R. Georgii \inst{3} \and
 A. von Kienlin\inst{3} \and
 G. Lichti\inst{3} \and
 V. Sch\"{o}nfelder\inst{3} \and
 A. Strong \inst{3} \and
 C. Wunderer\inst{3} \and
 C. Shrader\inst{4} \and
 S. Sturner\inst{4} \and
 B. Teegarden\inst{4} \and
 G. Weidenspointner\inst{4} \and
 J. Kiener\inst{5} \and
 M.-G. Porquet \inst{5} \and
 V. Tatischeff\inst{5} \and
 S. Crespin\inst{6} \and
 S. Joly\inst{6} \and
 Y. Andr\'{e}\inst{7} \and
 F. Sanchez\inst{8} \and
 P. Leleux \inst{9}}

\date{July 14, 2003; Accepted date}
\offprints{D. Atti\'e \\
(David.Attie@cea.fr)}
\institute{
 DSM/DAPNIA/Service d'Astrophysique, CEA Saclay, 91191 Gif-sur-Yvette, France, \and
 Centre d'Etudes Spatiales des Rayonnements, 9, avenue du Colonel Roche, 31028 Toulouse, France, \and
 Max-Planck-Institut f\"{u}r extraterrestrische Physik, Giessenbachstra\ss e, 85748 Garching, Germany, \and
 NASA/Goddard Space Flight Center, code 661, Greenbelt, MD 20771, USA, \and
 Centre de Spectrom\'etrie Nucl\'eaire et de Spectrom\'etrie de Masse, IN2P3-CNRS, 91405 Orsay Campus, France, \and
 CEA Bruy\`eres-le-Ch\^atel, BP 12 - 91680 Bruy\`eres-le-Ch\^atel, France, \and
 CNES/Centre Spatial de Toulouse 18 avenue E. Belin 31401 Toulouse, France, \and
 Universitat de Valencia, IFIC, Dr Moliner 50, 46100 Burjassot, Valencia, Spain, \and
 Univ. of Louvain, chemin du Cyclotron, B-1348 Louvain-La-Neuve, Belgium }

%%% ----------------------------------------------------------------
%%%       Abstract
%%% ----------------------------------------------------------------

\abstract{Three calibration campaigns of the spectrometer SPI have
been performed before launch in order to determine the instrument
characteristics, such as the effective detection area, the
spectral resolution and the angular resolution. Absolute
determination of the effective area has been obtained from
simulations and measurements. At 1 MeV, the effective area is 65
cm$^2$ for a point source on the optical axis, the spectral
resolution $\sim$ 2.3 keV. The angular resolution is better than
2.5\deg~and the source separation capability about 1\deg. Some
temperature dependant parameters will require permanent in-flight
calibration.

%%% ----------------------------------------------------------------
%%%       Key words
%%% ----------------------------------------------------------------

}

\maketitle \pretolerance=10000

%%% ----------------------------------------------------------------
%%%       1. Introduction
%%% ----------------------------------------------------------------

\section{Introduction}

During the SPI ground calibration at Bruy\`eres-Le-Ch\^atel (BLC),
low intensity radioactive sources were used at short distances (up
to 8 meters) for energy resolution,~camera efficiency and
homogeneity measurements. In addition, specific tests were
included using high intensity radioactive sources at 125 meters
for imaging performance measurements and photon beams generated
with a Van de Graaf accelerator were used for high energy
(E$_\gamma > 2.7$ MeV) calibration (\cite{Mandrou1},
\cite{Schanne}). Some additional runs with standard detectors were
necessary to understand the high energy lines. Different
GEANT\footnotemark[1] simulations which provided the Instrument
Response Function (IRF) (\cite{Sturner}) have been compared to
measurements.

The spectrometer SPI is described in \cite{Jean} \ and
\cite{Vedrenne}. In the text that follows the term \textit{camera}
refers to the Ge detector plane enclosed by the Anti Coincidence
System (ACS)\footnotemark[2] and the Plastic Scintillator
Anticoincidence sub-assembly (PSAC)\footnotemark[2]. The term
\textit{telescope} refers to the whole system of the camera with
the mask. The description of the camera and its events types (SE
and ME) is given in \cite{Vedrenne}.

\footnotetext[1]{{\scriptsize
http://wwwinfo.cern.ch/asdoc/geant\_html3/geantall.html}}
\footnotetext[2]{Spectrometer user manual is available at:
\newline {\scriptsize http://sigma-2.cesr.fr/spi/download/docs/mu/mu-5-2/:
\newline ACS: Vol. 1, p.25 , PSAC: Vol. 1, p.50}}

%%% ----------------------------------------------------------------
%%%       2. Calibration campaigns
%%% ----------------------------------------------------------------

\section{Calibration campaigns}

The SPI imaging capability has been tested using high intensity
sources ($^{241}$Am, $^{137}$Cs, $^{60}$Co, $^{24}$Na), located at
125~m from the telescope, outside the experiment hall through a
transparent window. For security reasons, the beam was strongly
collimated with a diameter of $\sim$2.5~m at the SPI position. In
order to ensure that the entire telescope was inside the beam, a
scanner using a NaI detector measured the vertical and horizontal
beam profiles. Before each SPI run, a standard Ge detector, whose
efficiency had been thoroughly calibrated, was used for 10 minutes
in order to obtain the real $\gamma$-ray flux entering SPI, thus
avoiding the need for any correction for the absorption within the
125 m air column.

The energy calibration and efficiency measurements were performed
with the mask removed so that the whole camera was illuminated
uniformly from a distance of about 8 m. A preliminary 6-day
monitoring campaign in the calibration hall with a Ge standard
detector demonstrated the absence of significant background
variations.

For the low energy measurements, eleven radioactive sources were
used in the range from 60 keV to 1.8 MeV. Sources emitting single
$\gamma$-ray lines or well separated lines were preferentially
selected. All source characteristics are listed in
 \mbox{Tab. \ref{tab2}}. In these data acquisitions, the sources were
placed at 8.533 m from the Ge detector plane. At this distance, we
can consider that each Ge detector is illuminated under the same
solid angle.

For the high energy range calibration, a high intensity proton
beam was directed onto a water-cooled thick $^{13}$C target (100
$\mu$g/cm$^2$), with SPI at 45\deg~from the beam axis. Two
resonances of the $^{13}$C(p,$\gamma$)$^{14}$N nuclear reaction at
E$_{p}$~=~550 and 1747 keV produce photons up to 9 MeV with
sufficient intensities. Relative line intensities at an angle of
45\deg~depend on the angular dependence of the $\gamma$-ray
emission. This effect has been measured taking into account all
absorption processes (\cite{Gros}). Since the intensity of the
proton beam on the target and the photon yield are not well known,
absolute efficiencies are not directly calculable. Thus we used a
two step process. The efficiencies obtained from accelerator
spectra were normalized to the 1638 keV line efficiency (Tab.
\ref{550keV} for 550 keV). The absolute efficiency at this energy
was calculated from the interpolation of the low-intensity source
efficiencies, the absolute efficiencies could then be derived for
the other accelerator lines.

A calibration phase with the \textit{INTEGRAL} satellite
completely integrated was performed at the ESA center of Noordwijk
(ESTEC). During the measurements \textit{INTEGRAL} was operated
vertically and irradiated by the sources previously used for the 8
m distance measurements at BLC permitting a comparison with the
BLC calibration results corrected for the mask.

%%% ----------------------------------------------------------------
%%%       3. Line fitting
%%% ----------------------------------------------------------------

\section{Line fitting}\label{sec:Fit}

The natural radioactivity background spectrum is subtracted from
the source spectrum. The photopeaks in the resulting spectrum are
fit by:
\begin{eqnarray}
F(x) & = &
\frac{N}{\sigma\sqrt{2\pi}}\int_{-\infty}^{+\infty}{{\rm
e}^{-\frac{1}{2}z^2}dz} + {\rm
erfc}\left(\frac{z}{\sqrt{2}}\right) +ax + b
\end{eqnarray}
where $z = \frac{|x-\mu|}{\sigma}$, and $x$ is the channel number
in the spectrum. The five parameters to be fit are $N$, $\mu$,
$\sigma$, $a$ and $b$, where $N$ is the amplitude of the Gaussian
profile, $\mu$ is the mean channel, $\sigma$ the Ge detector
resolution, $a$ and $b$ the coefficients of the linear function
modelling the residual background below the line. Inside the Ge
crystal, losses in the charge collection introduce a low energy
tail on the photopeak which is taken in account by the
complementary error function ${\emph{erfc}}$. Note that $\mu$ and
$\sigma$ are temperature dependent.

%%% ----------------------------------------------------------------
%%%       4. Energy restitution and energy resolution
%%% ----------------------------------------------------------------

\section{Energy restitution and energy resolution}\label{sec:Fit}

\begin{figure}[t]
  \begin{center}
    \epsfig{file=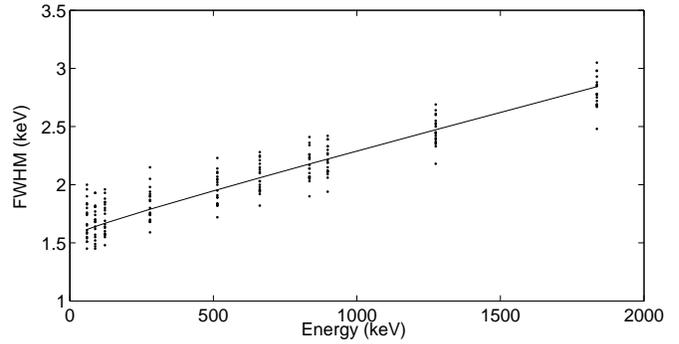,width=\columnwidth}
  \end{center}
    \vspace {-0.5cm}
    \caption{Spectral resolution (FWHM) for the flight model camera at 90 K.
    For each energy, the resolutions for the 19 individual detectors are plotted. The fit function is given by Eq. (\ref{Eq:FWHM}).} \label{FWHM}
\end{figure}

The spectral resolution and the energy-channel relation have been
measured during thermal tests for each Ge detector (\cite{Paul}).
They are temperature dependent. The mean resolution can be fit by
a quadratic function of~\emph{E}:
\begin{equation}\label{Eq:FWHM}
    {\rm FWHM} = \rm{F1} + \rm{F2} \times \sqrt{\emph{E}} + \rm{F3}\times \emph{E}
\end{equation}
For T = 90 K, F1 = 1.54, F2 = 4.6$\times$10$^{-3}$ and F3 =
6.0$\times$10$^{-4}$. At 1 MeV, FWHM $\sim$ 2.3 keV
(Fig.~\ref{FWHM}).

The energy-channel relations are almost linear for all detectors.
The variation of the peak position with temperature was $0.13 \,
\rm keV/ \rm K$ at $1.3 \, \rm MeV$ (measured in the temperature
range 93 K - 140 K). The relations obtained in the calibration
process are useless for in-flight data. The energy-channel
relations will change following the temperature differences.
%%%%%%%%%%%%%%%%%%%%%%%%
%%%%% Begin Figure %%%%%
%%%%%  End Figure  %%%%%
%%%%%%%%%%%%%%%%%%%%%%%%

%%% ----------------------------------------------------------------
%%%       5. Full-energy peak efficiency of the SPI telescope
%%% ----------------------------------------------------------------

%%%%%%%%%%%%%%%%%%%%%%%%
%%%%% Begin Figure %%%%%
\begin{figure}[h]
\epsfig{file=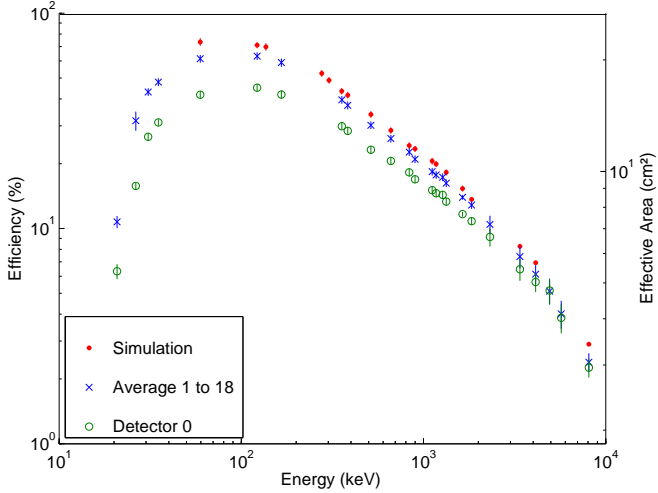,width=\columnwidth}
\caption{Full-energy peak efficiency for sources on the SPI
optical axis ($\alpha=0^\circ$), without mask. An average for
detectors 1 to 18 is shown; detector 0 is plotted separately (see
Sec. \ref{sec:Homo}). The measurements are compared to Monte-Carlo
simulations.} \label{Eff}
\end{figure}
%\vspace{-1.5cm}
%%%%%  End Figure  %%%%%
%%%%%%%%%%%%%%%%%%%%%%%%

\section{Full-energy peak efficiency of the SPI telescope}\label{sec:Eff}

To obtain the full-energy peak efficiency of the SPI telescope we
corrected the efficiency of the camera alone for the absorption of
photons by the mask. These results are compared to simulations.

%%%%%%%%%%%%%%%%%%%%%%%%
%%%%% Begin Figure %%%%%
\begin{figure*}[t]
    \vspace{1.4in}
 \begin{picture}(80,50)(0,-10)
    \epsfig{figure=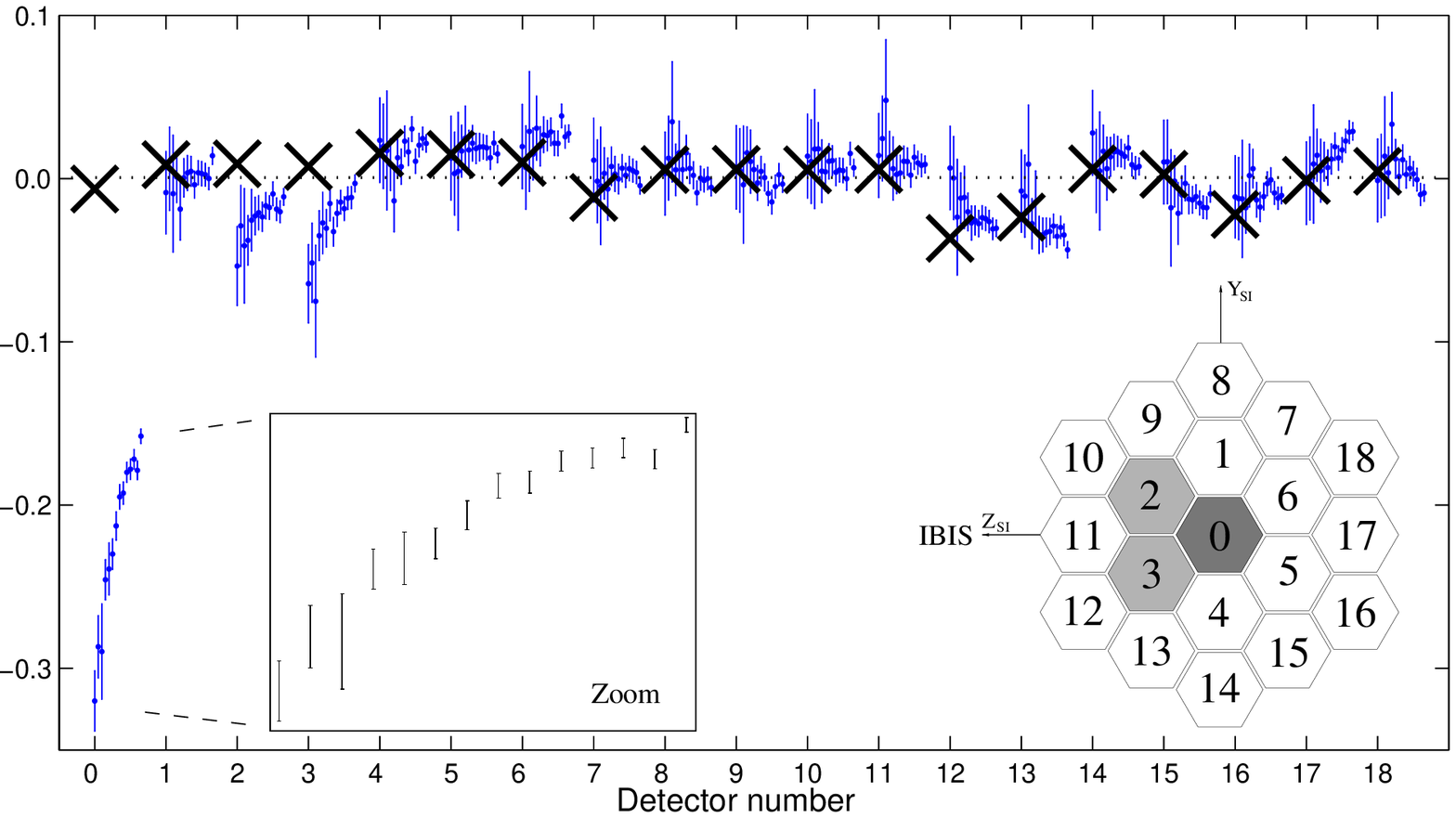,width=8.75cm}
 \end{picture}
 \begin{picture}(80,50)(-175,-20)
    \epsfig{figure=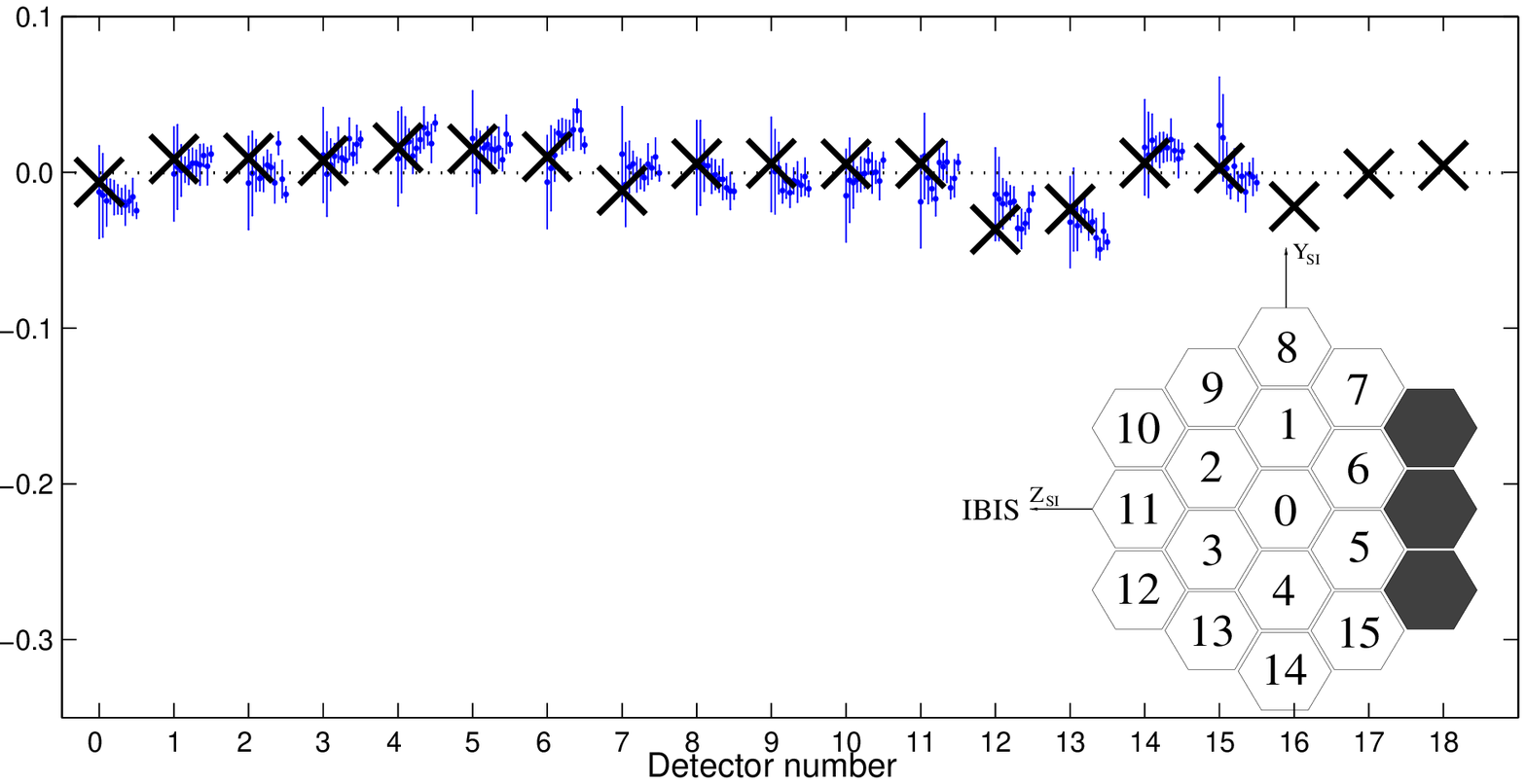,width=8.75cm}
 \end{picture}
    \vspace{-0.5cm}
    \caption{For each detector, a plot of the
    efficiency versus energy is shown (in each plot, the line energy
    increases from left to right according to the 14 energies used). The crosses
    represent the mass of the crystal. When the detector is not shadowed, the
    efficiency depends mainly on its mass. Left panel: at $0^\circ$,
    source on axis, the detectors 0,2 and 3 are partially shadowed by the
    PSAC-alignment device. Right panel: when the source is at
    $8^\circ$, detectors 16 to 18 are shadowed by the ACS. The
    alignment device is projected outside the detection plane.}
\label{All}
\end{figure*}
%%%%%  End Figure  %%%%%
%%%%%%%%%%%%%%%%%%%%%%%%

    \subsection{Individual detector efficiencies}\label{sec:Eff_noMask}

For a Ge detector $i$, and a $\gamma$-ray line at energy \emph{E}
produced by a source, the full-energy peak efficiency $\epsilon_i$
is defined by the ratio
\begin{equation}\label{Eq1}
    \epsilon_i = \frac{\phi_{\rm measured}}{\phi_{\rm incident}}
\end{equation}
where $\phi_{\rm measured}=\mathcal{A} / T_{\rm m}$. $\mathcal{A}$
is the number of  photons in the photopeak. We consider
$\mathcal{A}$ to be the integral of the Gaussian part of the
function $F(x)$ fitted to the background subtracted photopeak
($\mathcal{A}=\sqrt{2\pi} N \sigma$). $T_{\rm m}$ is the effective
measurement duration (\ie the total duration of the measurement
corrected for the dead time).

In the case of a source, the absolute flux $\phi_{\rm incident}$
at the detector plane is given by
\begin{equation}\label{Eq2}
    \phi_{\rm incident} = a_0 \times 2^{-\frac{T_{1}-T_{0}}{\tau}}
    \times {\cal B}r_E \times \Omega \times e^{-\frac{\mu_{\rm
    E}}{\rho}x}
\end{equation}
where $a_0$ is the source activity at the reference date $T_0$,
$T_1$ is the date of the measurement and $\tau$ the half-life of
the source, ${\cal B}r_E$ is the branching ratio of the line at
energy \emph{E}. The air transmission coefficient is computed
using the air mass attenuation coefficient $\mu_E$ at energy
\emph{E}, the air density $\rho$ and the distance $x$ between the
source and the Ge detection plane. $\Omega$ is the relative area
of the detector viewed from the source
%\begin{equation}\label{Eq2}
$\Omega~= \frac{A_{\rm Ge}}{4 \pi \times x^{2}}$
%\end{equation}
where $A_{\rm Ge}$ is the geometric area of a Ge detector,
$\langle A_{\rm Ge}\rangle = 26.75$ cm$^2$.

For the accelerator case, the intensities ${\cal I}r_E$ of the
lines in the accelerator spectrum, corrected for all absorption
effects, are relative to the 1638 keV line intensity
(Tab.~\ref{550keV}). The  efficiency is:
\begin{equation}\label{Eq3}
    \epsilon_{i}(E) = \frac{\phi_{\rm measured}(E)}{{\cal I}r_E \times \phi_{\rm measured}(1638)} \times \epsilon_{i}^{\rm{int}}(1638)
\end{equation}
where $\epsilon_i^{\rm{int}}(1638)$ is the detector efficiency at
1638 keV obtained from the interpolation of source data.

For detectors 1 to 18, efficiencies are comparable. The value
$\langle \widetilde{\epsilon} \rangle=\frac{1}{18}\sum_{i=1}^{~18}
\epsilon_i$  is representative of the efficiency of a single
detector. Efficiencies $\epsilon_0$  and $\langle
\widetilde{\epsilon} \rangle$ are compared to the GEANT simulation
average efficiency $\langle \widetilde{\epsilon} \rangle_{\rm
sim}$. $\langle \widetilde{\epsilon} \rangle_{\rm sim}$ is
$\sim$10~\% higher than $\langle \widetilde{\epsilon} \rangle$
(Fig. \ref{Eff}).

    \subsection{Homogeneity of the camera}\label{sec:Homo}

Using 14 of the 18 energies of Tab.~\ref{tab2}, we compared for
each detector i the efficiency homogeneity functions
$h_i(E)=\big[\frac{\epsilon_i-
\langle\widetilde{\epsilon}\rangle}{\langle\widetilde{\epsilon}\rangle}\big]_E$
to the corresponding mass homogeneity functions
$M_i~=~\frac{m_i-\langle m\rangle}{\langle m\rangle}$  where
$\langle m\rangle=\frac{\Sigma_i^n m_i}{n}$. In Fig. \ref{All} we
display the homogeneity functions and their energy dependence for
sources for a range of energies $E$ on the optical axis
($\alpha=0^\circ$) and for $\alpha=8^\circ$.

If $\alpha=0^\circ$, the efficiency of detector 0 seems to be
$\sim$~10~to~20 \% less than $\langle \widetilde{\epsilon}
\rangle$, and the deviation increases when the energy decreases.
This behaviour is the signature of absorption, which in the case
is due to an Hostaform plastic device inserted in the center of
the plastic anticoincidence scintillator (PSAC) for alignment
purposes. Detectors 2 and 3 are also affected. It was subsequently
found that during these calibration runs the sources were actually
slightly off-axis, causing the alignment device to partially
shadow detectors 2 and 3. This explains the slight attenuation
observed for them (Fig. \ref{All}, left panel).

If $\alpha \neq 0^\circ$, the alignment device is projected on one
or more other detector(s). During measurements at
$\alpha=8^\circ$, it is projected outside the camera, detector 0
shows a normal efficiency. The very low efficiency of detectors 16
to 18 is due to the shadow of the ACS (Fig. \ref{All}, right
panel).

%%%%%%%%%%%%%%%%%%%%%%%%
%%%%% Begin Figure %%%%%
\begin{figure}[!t]
    \epsfig{figure=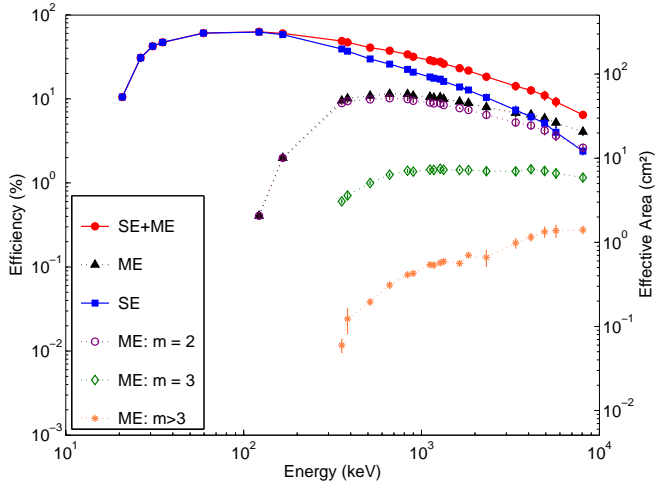,width=8.75cm}
   \caption{Full-energy peak efficiency for SE, ME and all events (SE+ME) and for ME
     with multiplicity \emph{m} = 2, 3 and $>3$, for the camera
     without mask. The effective area is the geometric
area multiplied by the efficiency.} \label{EffMESE}
\end{figure}
%%%%%  End Figure  %%%%%
%%%%%%%%%%%%%%%%%%%%%%%%

    \subsection{Full-energy efficiency of the camera for the Multiple Events (ME)}\label{sec:Eff_ME}

In the case of ME, the incoming photon cannot be associated with a
specific detector. So, the whole camera must be considered. We
constructed a spectrum for each calibration source and
multiplicity $m$ using events from the whole camera. We then fit
the lines in the background corrected spectra. The ME efficiency
for a multiplicity $m$, $\epsilon_{\rm ME}^{m}$, is defined as for
the SE in Eq. (\ref{Eq1}). We find that above $\sim$4 MeV, the
total ME efficiency is greater than the SE efficiency. The
contribution of different multiplicities is displayed in
Fig.~\ref{EffMESE}.

    \subsection{Full-energy peak efficiency of the telescope}

During the acquisition of the data used in the previous sections,
the mask was removed to let all detectors be illuminated by the
sources. Deriving the telescope efficiency from the camera and
detector efficiencies need to be corrected for the absorption of
photons by the open and closed elements of the mask. This
correction is evaluated for on-axis sources at infinite distance,
thus the rays are considered to arrive parallel on the mask.

The transparency of the 63 open mask pixels have been measured
individually with different radioactive sources from 17 keV  to
1.8 MeV (\cite{Sanchez}). Using these data, a mathematical model
was fit to reproduce the mask absorption for the open pixels and
especially for the central pixel, affected also by the alignment
device. Note that above 2 MeV the absorption values have been
extrapolated.

\begin{figure}[b]
    \epsfig{figure=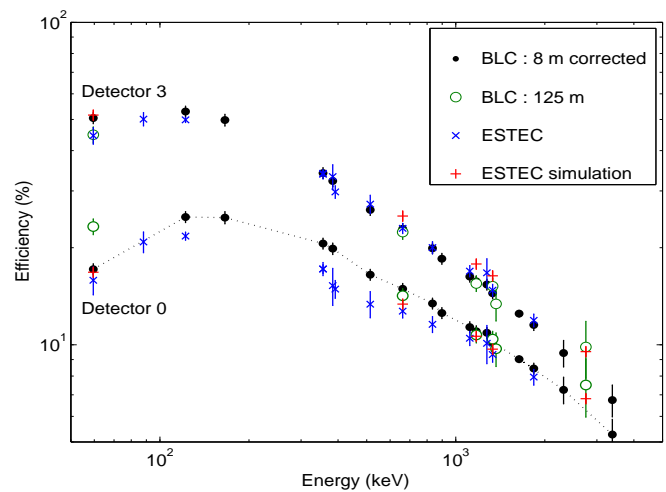,width=8.75cm,height=6.5cm}
    \caption{The on-axis, full-energy peak efficiencies of detector 0 and
    3 for SPI with its mask. The data are: 8 m BLC data without
    mask corrected for the  mask absorption ($\cdot$), 125 m BLC data with mask,
    ($\circ$) and  8 m ESTEC data with mask ($\times$). For detector 0, the disagreement
    between the curves shows the difficulty of  alignment
    device modelisation.}
    \label{BLCPLGC}
\end{figure}

The efficiencies computed from on-axis 8-meter source data
analyses for the detectors 0 and 3 were corrected for absorption
due to the mask support (Fig. \ref{BLCPLGC}). For detector 3,
these efficiencies are in good agreement with the efficiencies
obtained with the mask installed at ESTEC and at BLC using the
125-meter source data. Thus, for an unshadowed detector, the
correction method can be applied. For detector 0, when the source
is on axis, the presence of the alignment devices in the PSAC and
the mask complicates the efficiency calculation. Below 1 MeV, we
adjusted the correction (\cite{Sanchez}) to fit the BLC and ESTEC
measurements.

Let $M_{\circ}(E)$ be the mask support transmission for the
illuminated detectors, $M_{\bullet}(E)$ the mask element
transmission multiplied by the mask support transmission for the
shadowed ones. For SE, $M_{i}(E)$ is the transmission of the
element i of the mask at the energy $E$ for the detector $i$
($M_{i}(E) = M_{\circ}(E)$ or $M_{\bullet}(E)$). For SE we correct
the efficiency for each detector. For ME we correct the global
camera efficiency obtained in Sec~\ref{sec:Eff_ME} by  the global
mask absorption. The total effective area $\mathcal{A}_{\rm eff}$
of the telescope is then:
\begin{eqnarray*}
    \mathcal{A}_{\rm eff}  = \sum_{\rm i=0}^{\rm 18}
    \mathbf{A}_{i}^{SE}(E)\times
    M_{i}(E)
 \end{eqnarray*}
\begin{equation}
     ~~~~~~~~+  \mathbf{A}^{ME}(E) \times \Big[ \frac{\mathbf{A}_{\bullet}}{\mathbf{A}}
    M_{\bullet}(E) + \frac{\mathbf{A}_{\circ}}{\mathbf{A}}
    M_{\circ}(E)\Big]\
    \label{surfeff}
\end{equation}
$\mathbf{A}_{i}^{SE}(E)$ is the effective area of detector $i$ for
the SE, $\mathbf{A}^{ME}(E)$ is the camera effective area for the
ME. $\mathbf{A}_{\circ}$,~$\mathbf{A}_{\bullet}$~and~$\mathbf{A}$
are the total geometric area of the illuminated detectors, of the
shadowed detectors and of the whole camera. In the case of a
source on axis at infinity, $\mathbf{A}_{\circ}= 240.9~\rm cm^2$,
$\mathbf{A}_{\bullet}= 267.4~\rm cm^2$ and $\mathbf{A} =508.3$
cm$^2$.
%%%%%%%%%%%%%%%%%%%%%%%%
%%%%% Begin Figure %%%%%
%%%%%  End Figure  %%%%%
%%%%%%%%%%%%%%%%%%%%%%%%

For imaging, Eq. (\ref{surfeff}) is not fully valid for ME. In
this case, the number of the detector where the photon had its
first interaction can be known only with a probability $ p < 1 $.
The $(1-p)$ fraction is attributed to other pixels of the mask
(closed or open), and so the real ME efficiency is always less
than \textit{(ME~ camera~ efficiency)} $\times$ \textit{(mask~
absorption~ corrections)}.

\begin{figure}[t]
    \epsfig{file=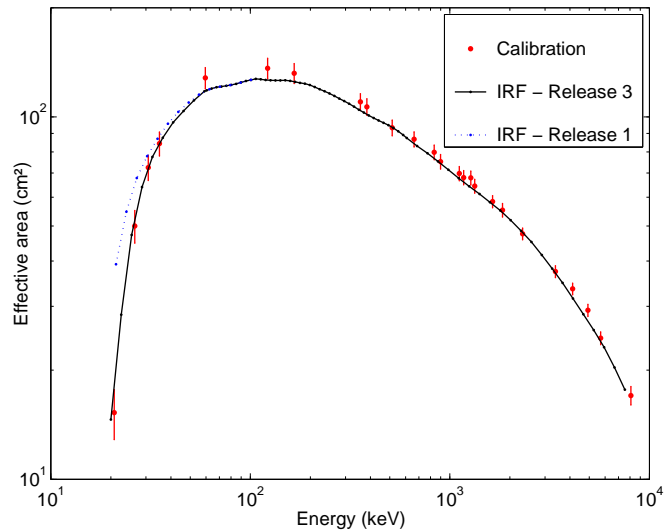,width=\columnwidth}
    \caption{Full-energy peak effective area of SPI telescope using
    all events (SE+ME). This curve is representative of the in flight
    effective area  for an on-axis source. We compare it to the response matrix before the
    launch (release~1,~nov~2002) and after the launch (release~3,~july~2003).}
\label{EffArea}
\vspace{-0.3cm}
\end{figure}

We have compared the measured effective areas of the SPI telescope
to those found in the SPI Imaging Response Files (IRFs), see
Fig.~\ref{EffArea}.  Here we have limited our comparison to the
on-axis full-energy peak effective areas.  The IRFs used in the
ISDC data analysis pipeline have been simulated using a
GEANT-based software package (\cite{Sturner}).
 Note that the version of the IRFs released in November 2002 have subsequently
been corrected at low energies using calibration data.  The total
effective area is about 125 cm$^2$ at 100 keV and 65 cm$^2$ at 1
MeV.

%%%%%%%%%%%%%%%%%%%%%%%%
%%%%% Begin Figure %%%%%
%%%%%  End Figure  %%%%%
%%%%%%%%%%%%%%%%%%%%%%%%

%%% ----------------------------------------------------------------
%%%       6. Imaging capabilities
%%% ----------------------------------------------------------------

\section{Imaging Capabilities}\label{sec:Imaging}

The long-distance source tests were designed to verify the
capabilities of the entire instrument in the imaging mode, the
response matrix derived  from Monte Carlo simulations, and the
performance of the instrument/software/response matrix
combination. For 125 m, the beam divergence was about $\pm 4'$ and
the angular size of the sources $\sim 1'$.

    \subsection{Angular resolution and Point Spread Function}

\begin{figure}
\vspace{0.2cm}\psfig{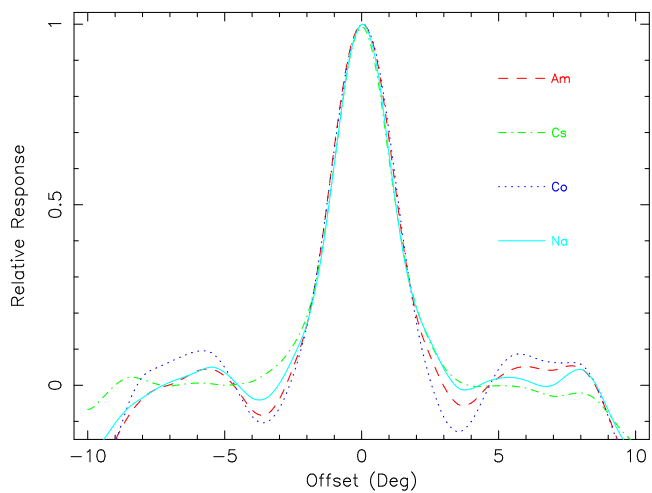}
      \caption[]{The PSF determined from
      measurements of 4 different sources at 125 m. Each curve is the mean of
      cross-sections through the response in two orthogonal directions; a
      background level has been subtracted from each  and the curves
      normalized to the same peak height. The mean FWHM is 2.55\deg. ($^{241}$Am: 60 keV; $^{137}$Cs: 662 keV; $^{60}$Co: 1332 keV; $^{24}$Na: 2754 keV)}
\label{PSF} \vspace{-0.4cm}
\end{figure}

The angular resolution and Point Spread Function (PSF) of an
instrument such as SPI is a function  not only of the
characteristics of the instrument but of the sequence of pointings
(the ``dithering pattern") used and the image reconstruction
technique adopted. Here we take the PSF to be the response in an
image obtained by correlation mapping (\cite{Skinner Connell}) to
a point source observed according to a specific dither pattern. We
then take as the angular resolution the full width at half maximum
(FWHM) of this response. Fig. \ref{PSF} shows some results.

%%%%%%%%%%%%%%%%%%%%%%%%
%%%%% Begin Figure %%%%%
%%%%%  End Figure  %%%%%
%%%%%%%%%%%%%%%%%%%%%%%%

Given that the hexagonal mask elements are 60 mm ``across flats"
and that the mask-to-detector distance is 1710 mm, the expected
angular resolution is 2.0\deg. But because of the finite detector
spatial resolution, the FWHM will be larger. Although the detector
pitch is equal to the mask element size, the detectors are
somewhat smaller (56 mm). This gives an expected FWHM of about
2.5\deg . The measured values are consistent with predictions. The
FWHM does not vary significantly with energy for the sources used
(59--2754~keV).

%%%%%%%%%%%%%%%%%%%%%%%%
%%%%% Begin Figure %%%%%
\begin{figure}[b]
\vspace{-0.4cm}
\hspace{0cm}\psfig{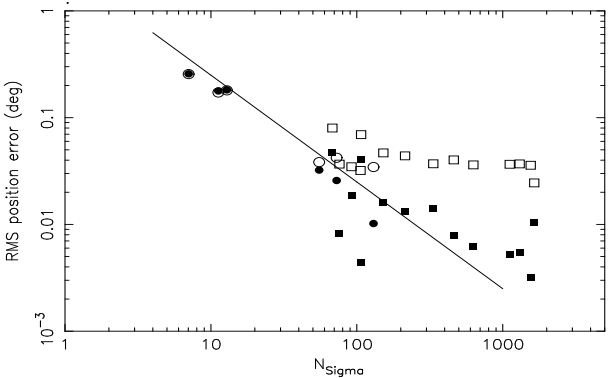}
      \caption[]{The error in the determination of
      a narrow line source position from analysis of
      calibration data. Squares: subsets of data,
      circles: mean of 10--20 trials with samples of
      random noise added. Open symbols: deviations from
      the nominal position, filled symbols: deviations from an
      assumed displaced source position. The line corresponds to
      2.5\deg$/N_\sigma$.}
\label{fig_pos}
\end{figure}
%%%%%  End Figure  %%%%%
%%%%%%%%%%%%%%%%%%%%%%%%

    \subsection{Single source localisation precision}

The source location accuracy depends on the signal-to-noise ratio,
$N_\sigma$, of the measurement as well as the angular resolution
of the instrument. Some analysis results of the analysis software
Spiros (\cite{Connell}, \cite{Skinner Connell}) are shown in Fig.
\ref{fig_pos}. The signal-to-noise ratio of the BLC data is
extremely high.  It is important to verify the performances with
values of $N_\sigma$ more representative of  flight values. To do
this, random subsets of events were taken. These subset were also
diluted by adding increasing amounts of Poisson-distributed noise.
We found that the position accuracy does not increase when the
signal-to-noise ratio exceeded 50-100$\sigma$. This suggests that
there are systematic effects, which limit the accuracy to $>2.5'$.
This could be due to uncertainties in the telescope stand
alignment, which are about that level. Assuming that the reference
axis was displaced from the source direction by a fixed 2.5$'$,
the residual errors suggest that the intrinsic limit of the
instrument may be about five times lower (filled symbols).

    \subsection{Source separation capability}
%%%%%%%%%%%%%%%%%%%%%%%%
%%%%% Begin Figure %%%%%
\begin{figure}[t]
\vspace{-0.5cm}
\hspace{0cm}\psfig{figure=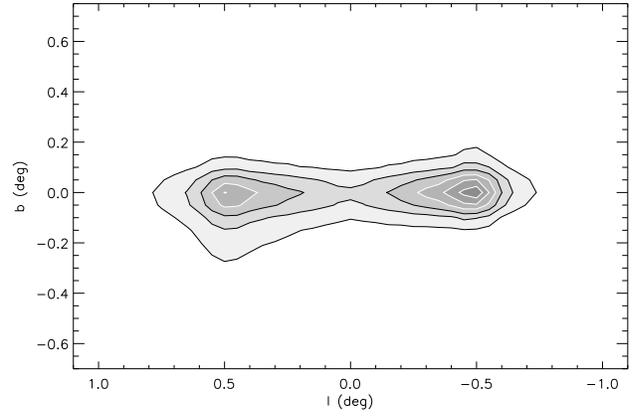,width=\columnwidth}
      \caption[]{Maximum entropy image in the 1173 keV  $^{60}$Co line
      shows the
      capability of the instrument to separate sources closer than
      the angular resolution, where the signal-to-noise ratio is good.
      The sources are separated by 1\deg.}
\label{fig_maxent} \vspace{-0.5cm}
\end{figure}
%%%%%  End Figure  %%%%%
%%%%%%%%%%%%%%%%%%%%%%%%

Even if the angular resolution is about 2.5\deg, one can
distinguish sources separated by less than this angle if data have
a good signal-to-noise ratio. The BLC calibration runs were
restricted to single sources, but it is possible to combine the
data from runs at different source angles to emulate data for
double sources. Fig. \ref{fig_maxent} shows Spiskymax
(\cite{Strong}) image for sources separated by 1\deg using the
1173 keV line of $^{60}$Co. The source are clearly separated, thus
showing that SPI, in the very high signal-to-noise regime of BLC,
is able to resolve sources at least as closely spaced as 1\deg .
With the lower signal-to-noise flight conditions the same
performance will not always be achieved.

%%% ----------------------------------------------------------------
%%%       7. Anti-Coincidence System performance
%%% ----------------------------------------------------------------

\section{Anti-Coincidence System performance}

\begin{figure}[h]
  \vspace {0.4cm}
  \epsfig{file=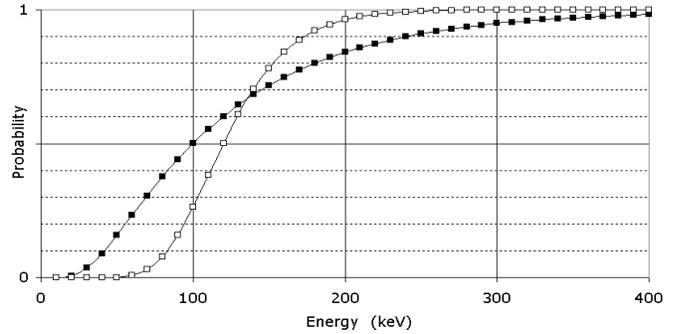,width=\columnwidth,height=4.5cm}
  \vspace {-0.4cm}
 \caption{For PMT$_1$(solid), $E_{\rm th}$ = 100 keV and $a = \sqrt{50}$, for PMT$_2$(open), 120 keV and $ \sqrt{10}$.
          These curves illustrate the large differences between ACS crystals and PMTs combinations.}
  \label{PMT ACS}
\end{figure}

A threshold calibration was performed using two radioactive
sources ($^{203}$Hg: 279 keV; $^{137}$Cs: 662 keV) placed close to
each of the 91 BGO crystals. The redundant cross-connections
between pairs of crystals, PMTs (photomultiplier tube) and
electronics give a broad threshold function, which can be
approximated by
\begin{equation}
    s(E,E_{\rm th}) = \frac{1}{\sigma(E) \times \sqrt{2\pi}} \times
    \int_{E_{\rm th}}^{+\infty}{e ^ {-\frac{(E~'-E) ^{ 2}}{2 \sigma ^{
    2}(E)}} dE~'}
\end{equation}
where s is the probability that a $\gamma$-ray with energy E
exceeds the threshold energy E$_{\rm th}$. A relation $\sigma (E)
= a \times \sqrt E $ has been assumed (Fig. \ref {PMT ACS}).
%%%%%%%%%%%%%%%%%%%%%%%%
%%%%% Begin Figure %%%%%
%%%%%  End Figure  %%%%%
%%%%%%%%%%%%%%%%%%%%%%%%

%%%%%%%%%%%%%%%%%%%%%%%%
%%%%% Begin Figure %%%%%
\begin{figure}[b]
    \epsfig{file=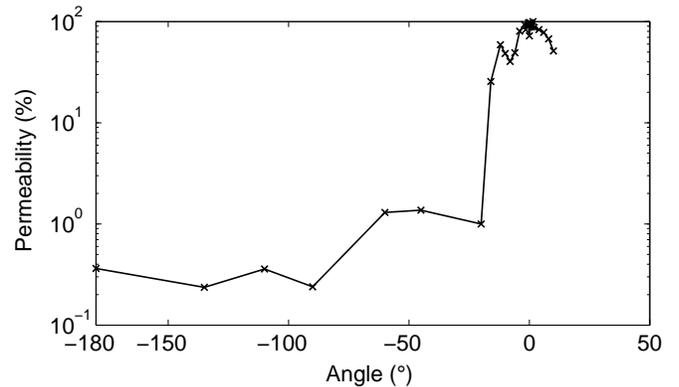,width=\columnwidth}
    \vspace {-0.4cm}
 \caption{ACS permeability as a function of the angle from the axis (BLC, $^{137}$Cs source)}
  \label{ACS}
\end{figure}
%%%%%  End Figure  %%%%%
%%%%%%%%%%%%%%%%%%%%%%%%

The light yield of BGO crystals varies with the temperature. This
has an significant effect on two main characteristics of the ACS:
\begin{itemize}
\item the self-veto effect is the rejection of true source counts
by the anticoincidence shield, due to Compton leakage from the
camera. It decreases when the ACS threshold energy raises as more
scattered events are accepted. For a detector located on the edge
of the camera, the influence is larger than for an inner one.
\item the veto-shield permeability is the fraction of
$\gamma$-rays, which go through the ACS shield undetected and then
hit a Ge detector. If a strong source is outside the field of
view, it will be \textit{imaged} through the structure of the veto
shield, which will act as a kind of mask (Fig. \ref{ACS}).
\end{itemize}

These two ACS effects will modify the in-flight sensitivity of the
telescope: variations of temperature on the orbit will affect the
Compton background under the lines and the dead time. Current
in-flight variations are about 10 K. They are permanently
monitored so that these effects could be further quantified.

%%% ----------------------------------------------------------------
%%%       8. SPI Background
%%% ----------------------------------------------------------------

\section{Background lines}\label{sec:Bkg}

We produced a catalogue of all the $\gamma$-ray lines detected by
SPI on the ground\footnotemark[1] \mbox{(Tab. \ref{BKG})} by
summing over all of the background data. Some never before
observed very-high energy lines were detected. These lines are
also visible in the \mbox{E$_p$ = 1747 keV} spectra, but not in
the \mbox{E$_p$ = 550 keV} ones. A careful study of BLC
environment excluded any kind of non-natural emission during our
background tests. A complementary test of $3 \times 10^{5}$ s
using an isolated standard Ge detector did not produce these
lines. So their observation in the SPI background spectra must be
related to the presence of low energy or thermal neutrons and the
Ge detector array structure of the camera.

During background measurements, the neutron flux comes mainly from
the spallation of SPI materials by cosmic rays. Thermal neutron
capture for Ge isotopes has been studied in details by
\cite{Islam}. They give the neutron separation energies: $S_n
(^{71}$Ge$) = 7415.9$ keV, $S_n(^{74}$Ge$) = 10196.3$ keV,
$S_n(^{75}$Ge$) = 6505.2$ keV. Lines are detected in SPI at these
energies, but the most striking feature is that the 10196 keV line
corresponds to a forbidden transition between the S$_n$ level and
the ground state.

A simulation of some of the numerous possible cascades in the
de-excitation of $^{74}$Ge nuclei showed that such a line can only
be observed due to the summation of different transitions of the
cascade by at least two detectors close to one another, \ie in
coincidence. On the other hand, the line corresponding to
$S_n(^{73}$Ge$) = 6782.9 keV$ is not observed, but a 6717 line is.
This line corresponds also to the summation of transitions, but
only down to the 67 keV level of the isomeric $ ^{73m}$Ge which
decays independently with a lifetime of 0.499 s.

The determination of line origins is necessary to understand the
SPI camera behaviour in space, where neutron and spallation
induced lines will affect the observations. The relative
intensities obtained in the ground calibration can then be used
when subtracting the background from astrophysical data.

\footnotetext[1]{\scriptsize http://www.edpsciences.org/}

%%% ----------------------------------------------------------------
%%%       9. Conclusion
%%% ----------------------------------------------------------------

\section{Conclusion}

SPI will be able to detect nuclear astrophysics lines and
continuum. The total effective area of SPI was found to be $>100$
cm$^2$ in the energy range 40 keV to 300 keV with a maximum of 136
cm$^2$ at 125 keV. At 1 MeV the resolution of the instrument is
$\sim$2.5 keV with an effective area of $\sim$65 cm$^2$. (At 511
keV: 90 cm$^2$; At 1.8 MeV: 52 cm$^2$).

The angular resolution of SPI was found to be roughly 2.5\deg and
it has been shown that SPI will be able to resolve sources with a
separation of 1\deg and probably less.

Using ground calibration data and GEANT simulations, we derived an
absolute calibration of the SPI effective area. We noticed the
temperature influence on ACS effects and energy restitution. For
these two points, we must underline that permanent in-flight
calibration is required. Neutron or spallation induced background
lines will be used as tracers to extract the background component
of some lines of astrophysical interest.

%%% ----------------------------------------------------------------
%%%       Acknowledgements
%%% ----------------------------------------------------------------

\begin{acknowledgements}

The authors would like to thank: %
C. Amoros, E. Andr\'e, A. Bauchet, M. Civitani, P. Clauss, I.
Deloncle, O. Grosjean, F. Hannachi, B. Horeau, C. Larigauderie,
J.-M. Lavigne, A. Lef\`{e}vre, O. Limousin, M. Mur, J. Paul, N. de
S\'er\'eville, J.-P. Thibault who took part in the shifts at BLC,
J.-P. Laurent (CISBIO) for the very tight logistics of high
intensity $^{24}$Na sources, the BLC Van de Graff team, M. A.
Clair (CNES), our project manager, R. Carli (ESA) ESTEC
calibration manager, D. Chambellan, B. Rattoni (DIMRI) for their
essential contribution to source preparation, P. Guichon and S.
Leray (CEA-Saclay), P. Bouisset, R. Gurrarian and E. Barker (IRSN
- Orsay) for fruitful discussion about nuclear physics.

\end{acknowledgements}

%%% ----------------------------------------------------------------
%%%       References
%%% ----------------------------------------------------------------

% ----------------------------------------------------------------
%       References
% ----------------------------------------------------------------

%%% ----------------------------------------------------------------
%%%       Annexes
%%% ----------------------------------------------------------------

% ----------------------------------------------------------------
%       Annexes
% ----------------------------------------------------------------

\begin{table*}[ht]
\begin{center}
\begin{tabular}{|c|c|c|c|c|c|c|c|}
  \hline
  E (keV)   &   Source      & $\epsilon_{\rm SE}$ (\%) & $\epsilon_{\rm ME}$ (\%)   & $T_{\rm ms}$ (\%) & $T_{\rm cp}$ (\%) & $A_{\rm eff}$ (cm$^2$) & $\Delta A_{\rm eff}$ (cm$^2$)    \\
  \hline
  \textbf{20.80}     &   $^{241}$Am  & 10.5  & 0.0   & 57.1  & 30.2  & \textbf{15.3}  & 1.4   \\
  \textbf{26.35}     &   $^{241}$Am  & 30.8  & 0.0   & 63.6  & 36.6  & \textbf{50.0}  & 4.2   \\
  \textbf{30.80}     &   $^{133}$Ba  & 42.2  & 0.0   & 67.2  & 40.8  & \textbf{72.5}  & 5.7   \\
  \textbf{35.07}     &   $^{133}$Ba  & 47.0  & 0.0   & 70.1  & 44.3  & \textbf{84.4 } & 6.4   \\
  \textbf{59.54}     &   $^{241}$Am  & 60.4  & 0.0   & 82.1  & 59.5  & \textbf{128.1} & 8.6   \\
  \textbf{122.06}    &   $^{57}$Co   & 62.2  & 0.4   & 83.7  & 66.1  & \textbf{136.2} & 8.9   \\
  \textbf{165.86}    &   $^{139}$Ce  & 58.1  & 2.0   & 84.4  & 68.9  & \textbf{131.9} & 8.4   \\
  \textbf{356.02}    &   $^{133}$Ba  & 39.1  & 9.5   & 86.1  & 75.9  & \textbf{110.0} & 5.9   \\
  \textbf{383.85}    &   $^{133}$Ba  & 36.9  & 10.1  & 86.3  & 76.6  & \textbf{106.6} & 5.6   \\
  \textbf{514.01}    &   $^{85}$Sr   & 29.8  & 10.9  & 86.9  & 79.2  & \textbf{ 93.3} & 4.7   \\
  \textbf{661.7}     &   $^{137}$Cs  & 25.9  & 11.5  & 87.5  & 81.6  & \textbf{ 86.7} & 4.2   \\
  \textbf{834.84}    &   $^{54}$Mn   & 22.4  & 11.4  & 88.0  & 83.7  & \textbf{ 79.8} & 3.8   \\
  \textbf{898.04}    &   $^{88}$Y    & 20.8  & 10.9  & 88.1  & 84.3  & \textbf{ 75.2} & 3.5   \\
  \textbf{1115.55}   &   $^{65}$Zn   & 18.2  & 10.6  & 88.6  & 86.3  & \textbf{ 69.8} & 3.2   \\
  \textbf{1173.24 }  &   $^{60}$Co   & 17.6  & 10.3  & 88.7  & 86.8  & \textbf{ 68.0} & 3.1   \\
  \textbf{1274.5}    &   $^{22}$Na   & 17.1  & 10.4  & 88.9  & 87.6  & \textbf{ 67.9} & 3.0   \\
  \textbf{1332.5}    &   $^{60}$Co   & 16.1  & 10.0  & 89.0  & 88.0  & \textbf{ 64.4} & 2.9   \\
  \textbf{1836.06}   &   $^{88}$Y    & 12.7  & 8.9   & 89.7  & 90.9  & \textbf{ 55.3} & 2.4   \\
  \hline
\end{tabular}
\begin{minipage}{0.8\linewidth}
\vspace{0.1cm} \caption{Energies of the radioactive sources used
at 8.533 meters and the associated efficiency calibration obtained
at Bruy\`eres-Le-Ch\^atel. The efficiencies camera for the SE and
ME is respectively given by $\epsilon_{\rm SE}$ and $\epsilon_{\rm
SE}$. The transparency of the mask support is $T_{\rm ms}$ and the
transparency of the central pixel is $T_{\rm cp}$. Using this
transmission factors and with an air density $\rho= 1.205 \times
10^{-3}$ g cm$^{-3}$ the effective area of the SPI telescope is
given by $A_{\rm eff}$ with the error $\Delta A_{\rm eff}$ in
cm$^2$.} \label{tab2}
\end{minipage}
\end{center}
\end{table*}

\vspace{0.2 cm}

\begin{table*}[ht]
    \begin{center}
        \begin{tabular}  {|c|c|c|c|c|c|c|c|c|} \hline
E$_{\gamma}$ (keV) & ${\cal I}_0$ & ${\cal I}r_{\rm E}$ & $\epsilon_{\rm SE}$ (\%) & $\epsilon_{\rm ME}$ (\%)   & $T_{\rm ms}$ (\%) & $T_{\rm dev}$ (\%) & $A_{\rm eff}$ (cm$^2$) & $\Delta A_{\rm eff}$ (cm$^2$) \\
\hline
\textbf{1637.9}  & \textbf{100}      & \textbf{100}      & 13.8&   9.3 &   89.5    &   89.8    &   \textbf{58.4}    &   2.3  \\
\textbf{2316}    & 139   $\pm$ 4.3   & 149   $\pm$ 4.6   & 10.4&   7.9 &   90.2    &   93.0     &   \textbf{47.6}    &   1.8  \\
\textbf{3383.8}  & 23.7  $\pm$ 0.8   & 27.0  $\pm$ 0.9   & 7.4 &   6.8 &   91.1    &   96.5     &   \textbf{37.4 }   &   1.4  \\
\textbf{4123}    & 102   $\pm$ 3.2   & 119   $\pm$ 3.7   & 6.1 &   6.5 &   91.5    &   98.3     &   \textbf{33.5}    &   1.2  \\
\textbf{4922.8}  & 16.3  $\pm$ 0.6   & 19.1  $\pm$ 0.7   & 5.1 &   5.9 &   91.9    &   99.9     &   \textbf{29.3}    &   1.1  \\
\textbf{5700.1}  & 12.8  $\pm$ 0.5   & 15.1  $\pm$ 0.6   & 4.0 &   5.2 &   92.2    &   99.9     &   \textbf{24.5}    &   1.0  \\
\textbf{8076}    & 627   $\pm$ 20    & 752   $\pm$ 24    & 2.4 &   4.1 &   93.0    &   99.9     &   \textbf{17.0}    &   0.7  \\
\hline
\end{tabular}
\begin{minipage}{0.8\linewidth}
 \caption {Strongest $\gamma$-ray lines of the 550 keV resonance of the
$^{13}$C(p,$\gamma$)$^{14}$N reaction. ${\cal I}_0$ is the
relative intensity emitted by the target under 45\deg, ${\cal
I}r_{\rm E}$ is the relative intensity at SPI position after
absorption in the target backing and holder, the cooling water,
8~m of air and the window separating SPI from the experimental
hall (normalized to the 1637.9~keV line). The 8076 keV line is
Doppler shifted from its nominal 8060 keV, as all lines coming
from the 8062 keV level. For the others values see Tab.
\ref{tab2}.} \label{550keV}
\end{minipage}
\end{center}
\end{table*}

\begin{table*}
\begin{center}
\tiny
\begin{tabular}{|ccccccc|}
  \hline
  Energy & Nuclide & Emission        & Half-life& Others gammas & Origin                    & Fluxes    \\
  (keV)  &         & probability (\%)&          & (keV) (\%)    &                           & (s$^{-1}$)\\
  \hline
 %30.0  &               &       &               &               &                           & -     \\
  46.5  & $^{210}$Pb    & 4.05  & 22.3 y        &               & $^{238}$U series (226Ra)  & 0.276 \\
  59.5  & $^{241}$Am    & 36.0  & 432.2 y       &               &                           & 0.094 \\
  63.3  & $^{234}$Th    & 4.5   & 24.1 d        &               & $^{238}$U series (226Ra)  & 0.365 \\
  66.7  & $^{73m}$Ge    & 100   & 0.499 s       &               & activation                & -     \\
  72.8  & Pb X-ray      & [100] &               & 75.0[100]     & fluorescence K$\alpha 1$          & 0.765 \\
  75.0  & Pb X-ray      & [60]  &               & 72.8[60]      & fluorescence K$\alpha 2$          & 0.822 \\
  75.0  & $^{208}$Tl    & 3.6   & 3.053 m       & 2614.6(99.8)  & fast neutron activation   & 0.822 \\
  84.8  & $^{208}$Tl    & 1.27  & 3.053 m       & 2614.6(99.8)  & fast neutron activation   & -     \\
  84.9  & Pb X-ray      & [35]  &               & 75.0[100]     & fluorescence K$\beta 1$          & -     \\
  87.3  & Pb X-ray      & [8.5] &               & 75.0[100]     & fluorescence K$\beta 2$          & -     \\
  93.3  & $^{228}$Ac    & 5.6   & 6.15 h        & 911.2(29.0)   & $^{232}$Th series         & 0.696 \\
 143.8  & $^{235}$U     & 10.9  & 7$\times10^8$ y& 185.7(57.2)  & natural                   & 0.132 \\
 162.7  & $^{235}$U     & 4.7   & 7$\times10^8$ y& 185.7(57.2)  & natural                   & 0.090 \\
 185.7  & $^{235}$U     & 57.2  & 7$\times10^8$ y& 143.8(10.9)  & natural                   & 0.835 \\
 186.1  & $^{226}$Ra    & 3.28  & 1600 y        &               & $^{238}$U series          & 0.835 \\
 198.3  & $^{73m}$Ge    & 100   & \emph{\textbf{0.499 s }}      &               & activation & 0.033 \\
 205.3  & $^{235}$U     & 4.7   & 7$\times10^8$ y&              & natural                   & 0.086 \\
 209.4  & $^{228}$Ac    & 4.1   & 6.15 h        & 911.2(29.0)   & $^{232}$Th series         & -     \\
 238.6  & $^{212}$Pb    & 43.6  & 10.6 h        & 300(3.34)     & $^{232}$Th series         & 0.753 \\
 240.8  & $^{224}$Ra    & 3.9   & 3.66 d        &               & $^{232}$Th series         & -     \\
 269.4  & $^{223}$Ra    & 13.6  &               &               & $^{235}$U series          & 0.014 \\
 270.3  & $^{228}$Ac    & 3.8   &               &               & $^{232}$Th series         & -     \\
 295.1  & $^{214}$Pb    & 19.2  & 26.8 m        & 3351.9(35.1)  & $^{238}$U series (226Ra)  & 0.176 \\
 300.0  & $^{212}$Pb    & 3.34  & 10.6 h        & 238.6(43.6)   & $^{232}$Th series         & 0.044 \\
 338.4  & $^{228}$Ac    & 12.4  & 6.13 h        & 911.2(29)     & $^{232}$Th series         & 0.124 \\
 351.9  & $^{214}$Pb    & 37.1  & 26.8 m        & 285.1(19.2)   & $^{238}$U series (226Ra)  & 0.323 \\
 409.6  & $^{228}$Ac    & 2.2   & 6.15 h        & 911.2(29.0)   & $^{232}$Th series         & -     \\
 432.8  & $^{212}$Bi    & 6.64  & 1.1 h         & 727.2(6.65)   & $^{232}$Th series         & -     \\
 463.1  & $^{228}$Ac    & 4.6   & 6.15 h        & 911.2(29.0)   & $^{232}$Th series         & 0.011 \\
 510.8  & $^{208}$Tl    & 22.8  & 3.053 m       & 2614.6(99.8)  & $^{232}$Th series         & -     \\
 511.0  &   many        &       &               &               & annihilation $e^+/e^-$        & -     \\
 569.6  &               &       &               &               & -                         & 0.125 \\
 583    & $^{208}$Tl    & 84.5  & 3.053 m       & 2614.6(99.8)  & $^{232}$Th series         & 0.100 \\
 609.3  & $^{214}$Bi    & 46.1  & 19.9 m        & 1120.3(15.0)  & $^{238}$U series (226Ra)  & 0.137 \\
 661.7  & $^{137}$Cs    & 85.1  & 30.17 y       &               & -                         & 0.016 \\
 665.5  & $^{214}$Bi    & 1.55  & 19.9 m        & 609.3(46.1)   & $^{238}$U series (226Ra)  & 0.015 \\
 726.8  & $^{228}$Ac    & 0.62  & 6.13 h        & 911.2(29)     & $^{232}$Th series         & -     \\
 727.2  & $^{212}$Bi    & 6.65  & 1.1 h         & 1620.7(1.51)  & $^{232}$Th series         & 0.085 \\
 755.3  & $^{228}$Ac    & 1.32  & 6.13 h        & 911.2(29)     & $^{232}$Th series         & 0.008 \\
 766.4  & $^{214}$Bi    & 4.83  & 19.9 m        & 609.3(46.1)   & fission ($^{95}$Zr)       & 0.066 \\
 766.4  & $^{234}$Pa    & 0.29  & 0.79 s        & 1001.0(0.83)  & $^{238}$U series          & 0.125 \\
 784.0  & $^{127}$Sb    & 14.5  & 3.85 d        & 685.7(35.3)   & fission                   & 0.032 \\
 794.8  & $^{228}$Ac    & 4.6   & 6.13 h        & 911.2(29)     & $^{232}$Th series         & 0.021 \\
 834.8  & $^{54}$Mn     & 99.98 & 312.3 d       &               & charged particle reaction & 0.023 \\
 860.6  & $^{208}$Tl    & 12.52 & 3.053 m       & 2614.6(99.8)  & $^{232}$Th series         & 0.025 \\
 904.3  & $^{214}$Bi    & 0.1   & 19.9 m        & 609.3(46.1)   & $^{238}$U series (226Ra)  & 0.318 \\
 904.3  & $^{228}$Ac    & 0.89  & 6.13 h        & 911.2(29)     & $^{232}$Th series         & 0.318 \\
 911.2  & $^{228}$Ac    & 29.0  & 6.13 h        & 969.0(17.4)   & $^{232}$Th series         & 0.320 \\
 934.1  & $^{214}$Bi    & 3.1   & 19.9 m        & 609.3(46.1)   & $^{238}$U series (226Ra)  & 0.014 \\
 964.6  & $^{228}$Ac    & 5.8   & 6.13 h        & 911.2(29)     & $^{232}$Th series         & 0.225 \\
 969    & $^{228}$Ac    & 17.4  & 6.13 h        & 911.2(29)     & $^{232}$Th series         & 0.259 \\
1001.0  & $^{234}$Pa    & 0.83  & 0.79 s        & 766.4(O.29)   & $^{238}$U series          & -     \\
1063.6  &               &       &               &               & -                         & 0.059 \\
1120.3  & $^{214}$Bi    & 15    & 19.9 m        & 609.3(46.1)   & $^{238}$U series (226Ra)  & 0.107 \\
1155.2  & $^{214}$Bi    & 1.7   & 19.9 m        & 609.3(46.1)   & $^{238}$U series (226Ra)  & 0.011 \\
1237    & $^{214}$Bi    & 5.96  & 19.9 m        & 609.3(46.1)   & $^{238}$U series (226Ra)  & 0.043 \\
1281    & $^{214}$Bi    & 1.48  & 19.9 m        & 609.3(46.1)   & $^{238}$U series (226Ra)  & 0.010 \\
1292    &               &       &               &               & -                         & 0.004 \\
1377    & $^{214}$Bi    & 4.15  & 19.9 m        & 609.3(46.1)   & $^{238}$U series (226Ra)  & 0.050 \\
1401.5  & $^{214}$Bi    & 1.39  & 19.9 m        & 609.3(46.1)   & $^{238}$U series (226Ra)  & 0.010 \\
1408    & $^{214}$Bi    & 2.51  & 19.9 m        & 609.3(46.1)   & $^{238}$U series (226Ra)  & 0.020 \\
1460.8  & $^{40}$K      & 10.67 & 1.28$\time10^9$ y & -         & natural                   & 2.056 \\
1492    &               &       &               &               & -                         & 0.015 \\
1496    & $^{228}$Ac    & 1.05  &               &               & $^{232}$Th series         & 0.009 \\
1508    &               &       &               &               &                           & 0.027 \\
1580    & $^{228}$Ac    & 0.71  &               &               & $^{232}$Th series         & 0.012 \\
1764.5  & $^{214}$Bi    & 16.07 &               &               & $^{238}$U series (226Ra)  & 0.229 \\
2614.4  & $^{208}$Tl    & 99.79 & 3.053 m       & 2614.6(99.8)  & $^{232}$Th series         & 0.545 \\
3197.0  & $^{208}$Tl    &       & 3.053 m       & 2614.6(99.8)  & $^{232}$Th series         & 0.011 \\
3475    & $^{208}$Tl    &       & 3.053 m       & 2614.6(99.8)  & $^{232}$Th series         &       \\
3708.1  & $^{208}$Tl    &       & 3.053 m       & 2614.6(99.8)  & $^{232}$Th series         & $<$0.001 \\
6129.2  & $^{16}$O      &       &               &               & -                         & $<$0.001 \\
6505.2  & $^{75}$Ge     &       &               &               & $^{74}$Ge(n,$\gamma$)     & $<$0.001 \\
6716.5  & $^{73}$Ge     &       &               &               & $^{72}$Ge(n,$\gamma$)     & $<$0.001 \\
7415.9  & $^{71}$Ge     &       &               &               & $^{70}$Ge(n,$\gamma$)     & 0.001 \\
7631.7  & $^{57}$Fe     &       &               &               & $^{56}$Fe(n,$\gamma$)     & $<$0.001 \\
7645.5  & $^{57}$Fe     &       &               &               & $^{56}$Fe(n,$\gamma$)     & $<$0.001 \\
10196   & $^{74}$Ge     &       &               &               & $^{73}$Ge(n,$\gamma$)     & $<$0.001 \\
\hline
\end{tabular}
\begin{minipage}{0.8\linewidth}
\vspace{0.1cm} \normalsize \caption{Background environment lines
observed by SPI in absence of sources.} \label{BKG}
\end{minipage}
\end{center}
\end{table*}

%%%%%%%%%%%%%%
\end{document}